\documentclass{PoS}

\newcommand\beq{\begin{equation}}
\newcommand\eeq{\end{equation}}
\newcommand\beqa{\begin{eqnarray}}
\newcommand\eeqa{\end{eqnarray}}

\title{Hybrid Quark Stars With Strong Magnetic Field}

\ShortTitle{Hybrid quark stars}

\author{\speaker{Toshitaka Tatsumi}\\
        Department of Physics,Kyoto University, Kyoto 606-8502, Japan\\
        E-mail: \email{tatsumi@ruby.scphys.kyoto-u.ac.jp}}

\author{Hajime Sotani\\
        Division of Theoretical Astronomy, National Astronomical Observatory of Japan, 2-21-1 Osawa, Mitaka, Tokyo 181-8588, Japan\\
       E-mail: \email{sotani@yukawa.kyoto-u.ac.jp}}

\abstract{Discovery of huge magnetic field in magnetars has stimulated a renewed interest about the magnetic field and physics of compact stars, where microphysics such as QED or QCD may play active parts.
 Here we discuss the equation of state (EOS) of quark matter in the core of compact stars by taking into account the strong magnetic field. We show that quark EOS becomes very stiff in the presence of the strong magnetic field, and becomes stiffest under the causality condition beyond the threshold strength of $B_c\sim O(10^{19})$ G. This is because quarks make the Landau levels in the presence of the magnetic field and thereby only the lowest Landau level is occupied in the extreme case beyond $B_c$. Thus quarks can freely move along the magnetic field with localization in the perpendicular plane, which resembles the quasi-one dimensional systems and gives rise to a stiff EOS.  Consequently, we may easily produce high-mass stars beyond two solar mass.

As another interesting possibility, we discuss the appearance of the third family of compact stars, succeeding white dwarfs and neutron stars, before collapsing into black holes. We demonstrate an example, which is specified by a discontinuous increase of the adiabatic index at the hadron-quark phase transition. Such new family may affect the supernova explosions or the gravitational wave emitted from the neutron star mergers.}

\FullConference{The 26th International Nuclear Physics Conference\\
		11-16 September, 2016\\
		Adelaide, Australia}

\begin{document}

\section{Introduction}

There are some important issues in compact star phyasics, related to modern nuclear physics. One is the origin of the strong magnetic field in pulsars and magnetars \cite{cha}. Since it may be rather difficult to attribute  it to the astrophysical origin, a microscopic origin caused by the properties of nuclear matter might be interesting \cite{tat,yos}. Secondly, recently observed massive neutron stars with beyond 2$M_\odot$ have put a severe constraint on the equation of state (EOS) given by nuclear physics \cite{dem,and}, where hyperon or quark degree of freedom is expected to appear. Finally emergence of quark degree of freedom may be interesting in the core region of compact stars. More interestingly, one may expect the existence of quark stars as a third family of compact stars. One may give answers to these issues by making use of recent theoretical and observational knowledge.

On the other hand, the effect of the strong magnetic field on the nuclear EOS becomes one of the main subjects, stimulated by the recent discovery of magnetars \cite{mag}. Its strength amounts to $O(10^{15})$G at the surface, which corresponds to $O(10)$MeV as the energy scale. If the density-dependent variation of the magnetic field is taken into account, it may be much amplified in the core region of compact stars. In this case the energy scale becomes comparable with the strong-interaction energy scale, and the effect of magnetic field 
becomes remarkable. 

In this talk we discuss a possibility of massive compact stars with beyond 2$M_\odot$ in the presence of huge magnetic field \cite{sot1}. After that we briefly discuss a possibility of third family of compact stars using the same idea \cite{sot2}. 

\section{Model}

We assume almost free quark matter in the presence of the strong magnetic field in the core region surrounded by  hadronic matter in the low density region. 

There are many studies of EOS with the magnetic field \cite{men,hua,cas}. Taking the magnetic field $B$ along the $z$- axis, the Landau quantization leads to the quark single-particle energy,
\beq
E(n,k_z)=\left(k_z^2+\left(\sqrt{m_q^2+2\nu\left|qB\right|}-s\mu_N\kappa B\right)^2\right)^{1/2},
\eeq
with $\nu=n+1/2-{\rm sign}(q)s/2$ and the anomalous magnetic moment $\kappa$. In the following we discard $\kappa$, since it is a minor effect. Accordingly the wave function is localized in the $x-y$ plane, while still uniform along the $z$- axis. Thus quarks produce the one dimensional Fermi sea for each Landau level $n$. 

There are two remarks. Different from neutron matter, all quarks are charged and are efficiently subjected to the magnetic field. Different from charged hadrons such as hyperons, quark mass is very small, compared with $\sqrt{qB}$. It is also very small compared with the Fermi energy, so that quark matter is ultra-relativistic.
So, we can expect that the effect of the magnetic field is pronounced in quark matter. In the following we consider massless $u,d,s$ quark matter.

As the magnetic field becomes stronger, the fewer Landau levels are occupied from the lowest one (LLL). In the extreme case, only LLL is occupied beyond the critical strength $B_c$. Number density for each flavor $f$ is given by    
\beq
n_f=3\frac{\left|e_f B\right|}{2\pi^2}p_{fF}
\eeq
for each Landau level ($m_q\sim 0$). Hence the critical magnetic field is density dependent and given by 
\beq
eB_c=\left(6\pi^4\right)^{1/3}n_b^{2/3},
\eeq
with baryon number density $n_b=\sum_f n_f$, wich amounts to $O(10^{19})$ G. Assuming the virial theorem, the maximum magnetic field strength can be $\sim 10^{18}$ G for the neutron star with $R=10$ km and $M=1.4M_\odot$ \cite{lai}. We remark that this maximum field strength is derived on the assumption that the  field strength inside the star is almost constant and based on Newtonian gravity. Thus, for the case of neutron stars whose surface magnetic fields are not so strong, the magnetic field in the vicinity of the stellar center might be significantly stronger than $\sim 10^{18}$ G consistently to the virial theorem. In fact, the observed surface strength of magnetic field of white dwarf is at most in the range of $10^{6}-10^8$ G, while the central field strength in the order of $10^{12}$ G can be theoretically constructed \cite{ost}. Considering an analogy between magnetic fields inside neutron stars and white dwarfs, the central field strength of neutron stars may amount to $\sim 10^{19}$ G.

\begin{figure}
\begin{center}
\includegraphics[scale=0.5]{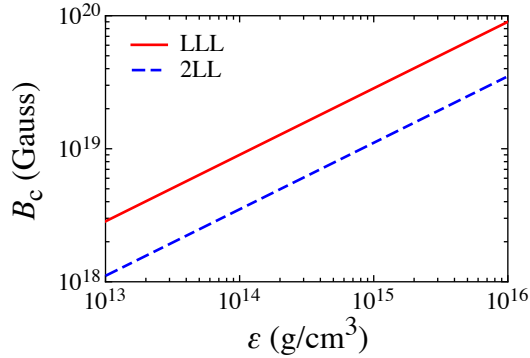} 
\end{center}
\caption{
The critical magnetic field strengths so that the quark matter settles only in the lowest Landau level (LLL) and up to the 2nd Landau level (2LL) are shown as a function of the total energy density $\varepsilon$.
}
\label{fig:Bc}
\end{figure}

Then the energy density is given as
\beq
\varepsilon_f=\frac{3\left|e_f B\right|}{4\pi^2}p_{fF}^2, 
\eeq
for each flavor $f$. Using Eqs.~(2.2),(2.4), we can write down EOS of quark matter with magnetic field beyond $B_c$,
\beqa
P&=&\frac{5\pi^2}{2eB}n_b^2-B_{\rm bag}\nonumber\\
  &=&\epsilon-2B_{\rm bag},
\eeqa
where $\varepsilon$ is the total energy density, $\varepsilon=5\pi^2 n_b^2/(2eB)+B_{\rm bag}$ and $B_{\rm bag}$ is the bag constant within the MIT bag model. Note that EOS  $P(\varepsilon)$ is independent of $B$, while both pressure and energy include the effect of the magnetic field.
There is another specific features of our EOS.  It gives the stiffest EOS, for which the sound velocity approaches the light velocity, $c_s^2\rightarrow 1$ (causality limit) 
\footnote{Similar idea has been applied for white dwarfs to discuss  possibility of super-Chandrasekhar-mass white dwarfs \cite{muk}.}
. On the other hand, $c_s^2\rightarrow 1/3$  in the absence of the magnetic field. This feature is similar to the repulsive effect given by the massive vector meson in hadronic matter or the effective vector interaction in the NJL type models \cite{zel,mas}.

Connecting the quark EOS with the hadronic one at low density region, we obtain full EOS's (Fig.~2). If we connect two EOS by a thermodynamically consistent way, we must carefully treat, e.g., the mixed phase, which leads to the Maxwell construction or the Gibbs construction. We demonstrate an example in Fig.~2 as "Maxwell" \cite{mar} without the magnetic field for comparison. For other cases we use Eq.~(2.5) for quark EOS for various values of $B_{\rm bag}$ and simply connect two EOS at some densities.

\begin{figure}
\begin{center}
\includegraphics[scale=0.5]{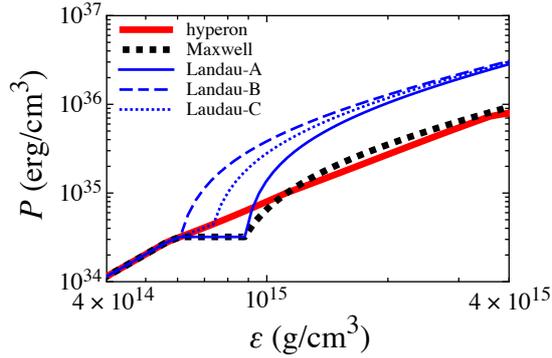} 
\end{center}
\caption{
Examples of EOS, where we use the hadronic EOS by the Brueckner-Hatree-Fock calculation with hyperons \cite{mar}.
}
\label{fig:EOS}
\end{figure}

\begin{figure}
\begin{center}
\includegraphics[scale=0.5]{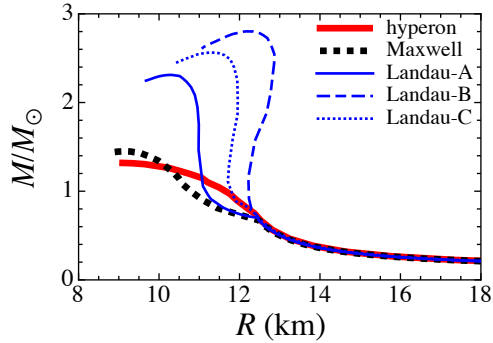} 
\end{center}
\caption{Mass-radius relations of hybrid stars constructed with several EOSs.}
\label{fig:MR}
\end{figure}

Solving the TOV equation, we have the mass-radius relation (Fig. \ref{fig:MR}).

Since the quark EOS becomes stiffer than the hadronic EOS in the high density region, hybrid stars with quark core can be more massive than those without quark core. However, the maximum mass of hybrid stars without the magnetic field is still too small to explain the observed mass of $2M_\odot$. Consequently, the effect of the magnetic field increases the maximum mass up to well above $2M_\odot$. It should be interesting to see that the ratio of the radius $R$ and that of quark core $R_Q$, $R_Q/R$,  amounts to $0.8$ for $M=2M_\odot$, which is quite different from the massive hybrid stars suggested up to now, where the quark core is usually very tiny \cite{sot1}.

\section{Possibility of the third family of compact stars}

We know that there is a sequence of compact stars as the central density increases and the radius becomes small: white dwarfs and neutron stars exist before collapsing into black holes. Usually stars become gravitationally unstable beyond the maximum central density of neutron stars and collapse into black holes. Then, one may ask a possibility of the third family of compact stars, which should be stabilized by new effects than ever considered.
There has been discussed such possibility by considering the phase transition to pion condensation, hyperon mixing or quark matter. However, it is impossible to make massive stars beyond $2M_\odot$ due to the softening of EOS caused by the phase transitions \cite{kem,sch}.

\begin{figure}
\begin{center}
\includegraphics[scale=0.5]{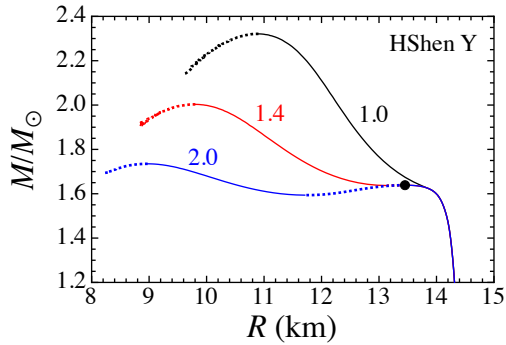}
\end{center}
\caption{
Mass-radius relations constructed with the HShen Y EOS connected to the quark EOS at $\varepsilon_c=1.0$, $1.4$, and $2.0\times 10^{15}$ g/cm$^3$. In the figure, the solid and dotted lines correspond to the stable and unstable equilibrium configurations, respectively.  For reference, the maximum mass of neutron stars expected with the HShen Y EOS denotes by the filled circle in the figure.
}
\label{fig:MR-HShenY}
\end{figure}

We discuss this issue by using our model. Taking an EOS given by Shen et al. as a typical hadronic EOS this time \cite{she}, we construct the EOS for hybrid stars by simply connecting two EOS in the similar way to the above discussion,
\beq
p(\varepsilon)=\left\{
  \begin{array}{@{\,}cc@{\,}}
     p_H(\varepsilon) & \varepsilon < \varepsilon_c \\
     p_Q(\varepsilon) & \varepsilon \ge \varepsilon_c
  \end{array}\right. ,
  \label{hqeos}
\eeq
where $\varepsilon_c$ is the transition density from hadronic pahse to quark one. We demonstrate some examples of the third family of compact stars beyond neutron stars. Solving the TOV equation again, we have the mass-radius relation (Fig. \ref{fig:MR-HShenY}). One can then observe the third family of compact objects for the cases of $\varepsilon_c=1.4$ and $2.0\times 10^{15}$ g/cm$^3$, for which the branch of hybrid stars is clearly separarted from the neutron stars by the gravitationally unstable region. Since the pressure of quark matter mainly supports such stars against gravitation, we may call them "quark stars", distinct from hybrid stars with small quark core.

\begin{figure}
\begin{center}
\includegraphics[scale=0.5]{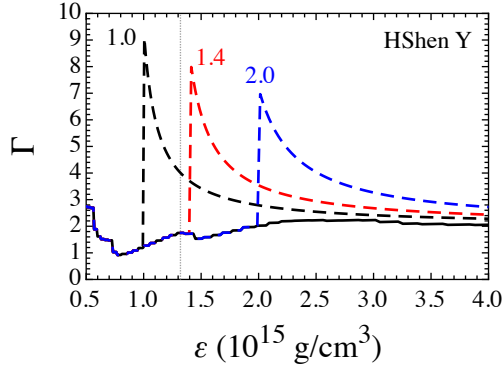}
\end{center}
\caption{With the HShen Y EOSs, the adiabatic index is shown as a function of the energy density, where the solid and broken lines correspond to the results for the stellar models without and with the phase transition into quark matter. The labels in the figure denote the transition density $\varepsilon_c$ in the unit of $10^{15}$ g/cm$^3$. 
For reference, the vertical line denotes the central density with which the mass of neutron star constructed with hadronic EOS becomes maximum.
}
\label{ad}
\end{figure}

One may get some hints about how massive quark stars become possible in terms of the adiabatic index, 
\begin{equation}
  \Gamma = \frac{p+\varepsilon}{p}\left(\frac{dp}{d\varepsilon}\right).  \label{eq:gamma}
\end{equation}
In Fig. \ref{ad} is shown the change of the adiabatic index inside quark stars.
 It's value asymptotically approaches to 2 in the high-density limit to satisfy the necessary condition for the gravitational stability of superdense stars,
\begin{equation}
  \Gamma>\frac{4}{3}\left(1+K\frac{R_s}{R}\right),
  \label{cond1}
\end{equation} 
where the second term is the general relativistic correction: $R$ is the stellar radius, $R_s=2MG/c^2$ is the Schwarzschild radius, $K$ is of order unity, and $M$ is the stellar mass. 
Besides this condition we can see the sharp increase of $\Gamma$ at the transition point: once quark matter appears during the gravitational collapsing, it rapidly exerts high pressure to interrupt collapsing. These features should be very important. However, comparing three curves in Fig. \ref{fig:MR-HShenY}
we notice the transition density, which should exceed the central density of maximum-mass neutron stars,  should be rather small to get massive stars. Otherwise, the graviational effect overwelms the effect of the magnetic field.  

\section{Summary and concluding remarks}

The compact stars have a strong magnetic field. Maybe, one should consider the effects of such strong magnetic field on the structure of compact objects. We particularly focus on the hybrid stars in this talk. We estimate the critical magnetic field strength that only the lowest Landau level is occupied, which is shown to be $B\sim O(10^{19})$ G. We also find that the equation of state (EOS) in the phase of the lowest Landau level can be expressed as $P=\varepsilon-2B_{bag}$ independently of the magnetic field strength. We remark that this is the limit of a stiff EOS, i.e., the sound velocity becomes equal to the light velocity. Usually the maximum mass of hybrid stars comes to be smaller than the observed maximum mass, i.e., $2M_\odot$, because the introduction of quark matter makes the EOS soft. However, we successfully construct the massive hybrid star models, owing to the effect of the Landau levels in the quark phase, which can be larger than $2M_\odot$. Furthermore, in order to examine the qualitative behavior of hybrid star models, we simply consider three different connections of quark matter with the hadronic matter. As a result, we find that the stellar model constructed with EOS connected to the hadronic matter at the lower energy density can realize more massive stellar model with smaller central density.   

Next, we have considered the possibility for existence of the third family of compact objects, which are stable equilibrium configurations more compact than neutron stars. It is not quite sure whether the stable objects more compact than neutron stars can exist or not. To construct such compact objects,  we have considered the magnetic field inside hybrid stars, whose strength and geometry are not fixed observationally. If the magnetic fields inside the star would be so strong that quarks settle only in the lowest Landau level, the quark EOS becomes quite stiff. 

Then, we have shown that the third family of compact objects can exist, even though the relevant parameter space is not so large and the hadronic EOS might be restricted, considering the observational constraint on the stellar mass such as $2M_\odot$. If such stars exist, they are (hybrid) quark stars composed of primarily quarks. By way of the numerical examinations, we have found that, for constructing the third family of compact objects, the hadronic EOSs could be favored, with which the maximum mass of neutron star should be small with a small central density and a large radius. That is, considering the possibility for the existence of third family of compact objects, the hadronic EOSs including hyperons may not be ruled out, even if the maximum mass of neutron star constructed with such soft hadronic EOSs is predicted to be less than $2M_\odot$. We additionally remark that, if the third family exists, the compact objects whose radii are smaller than neutron stars with the same mass can exist even in the narrow mass range, i.e., the existence of twin stars. Furthermore, if compact objects in the third family is formed in supernovae with larger masses than the maximum mass of neutron stars, one might expect to observe the second bounce when the central density reaches the transition density where quark matter appears, which is similar to the usual bounce in supernovae arising when the central density reaches the nuclear saturation density. As another possibility, one might distinguish compact objects in the third family from usual neutron stars by observing the thermal evolution, because the cooling history could strongly depend on the interior compositions.


We have seen that EOS is strongly stiffened after the deconfinement transition due to the strong magnetic field, and the adiabatic index becomes larger than the critical value of the gravitational stability, $4(1+KR_s/R)/3$. However, only the large adiabatic index in EOS is not sufficient for the possible existence of the third family. The transition density must be not so large, and at the same time the adiabatic index following the phase transition must increase enough to large: a combination of both conditions can produce the third family. These observations may partially support the conjecture suggested in \cite{ger}. Additionally, we have neglected the magnetic pressure for constructing the stellar models. At least, the poloidal magnetic fields can increase the maximum mass of the stars, which may be an advantage for realizing the third family of compact objects. 

\section*{Acknowledgement}

This work was supported in part by Grants-in-Aid for Scientific Research on Innovative Areas through 
No.\ 24105008 and No.\ 15H00843 provided by MEXT, and by Grant-in-Aid for Young Scientists (B) through No.\ 26800133 provided by JSPS.

\end{document}